\documentclass[twocolumn,prl,aps,floats,showpacs]{revtex4}
\usepackage{graphicx}

\begin{document}
\newcommand{\re}{\,{\rm Re}\,}
\newcommand{\im}{\,{\rm Im}\,}
\newcommand{\tr}{\,{\rm tr}\,}

\title{Quantum noise and self-sustained radiation of ${\cal PT}$-symmetric systems}

\author{Henning Schomerus}
\affiliation{Department of Physics, Lancaster University, Lancaster, LA1 4YB,
United Kingdom}

\date{\today}
\begin{abstract}
The observation that ${\cal PT}$-symmetric  Hamiltonians can have real-valued
energy levels even if they are non-Hermitian has triggered intense activities, with experiments in particular focusing on optical
systems, where Hermiticity can be broken by absorption and amplification. For
classical waves, absorption and amplification are related by time-reversal
symmetry. This work shows that microreversibility-breaking quantum noise
turns ${\cal PT}$-symmetric systems
into self-sustained sources of radiation, which distinguishes them from ordinary, Hermitian quantum systems.
\end{abstract}

\maketitle

A common factor in quantum systems with a non-Hermitian Hamiltonian is the non-conservation of particle number, either because the system is leaky, or because there is loss or gain in an absorbing or amplifying medium. Ignoring nonlinear effects
such as the feedback in a laser, such systems ordinarily do not possess stationary states; instead, they only support decaying quasibound states with complex energy, where the imaginary part ${\rm Im }\,E= -1/2\tau$ (setting $\hbar \equiv 1$) accounts for particle loss with decay rate  $1/\tau$
(particle gain is associated to a negative decay rate).
A notable exception are non-Hermitian systems with loss and gain combined such that they are invariant under joint parity (${\cal P}$) and time-reversal  (${\cal T}$) symmetry \cite{bender1998}.
If there is no leakage, these ${\cal PT}$-symmetric  systems generically possess a set of real-valued energy levels, as well as complex-conjugate pairs of complex energy levels. Systems with entirely real spectrum define a consistent unitary extension of quantum mechanics  \cite{bender2002,mostafazadeh}.
This observation has led to intense research efforts delivering a new theoretical perspective on systems as varied as quantum field theories and complex
crystals \cite{bender2007a}, while
experimental realizations in particular focus on optical systems where Hermiticity can be violated by absorption and amplification  \cite{christodoulides}.

For classical waves, amplification and absorption are strictly related by
time reversal. The existence of stationary states with real
energy can therefore be seen as a consequence of the balance of amplification
and absorption in parity-related regions of a ${\cal PT}$-symmetric system. At
the heart of absorption and amplification, however, are noisy microscopic
quantum processes (such as spontaneous and stimulated emission events, and stimulated absorption events)
which effectively break time-reversal symmetry
\cite{einstein}.
The objective of this Letter is to show that the effects of this quantum noise distinguish  ${\cal
PT}$-symmetric systems from Hermitian quantum systems, and indeed suggest
an alternative interpretation of the physics behind non-Hermitian ${\cal
PT}$ symmetry: (i) Accounting for quantum noise,  ${\cal PT}$-symmetric
systems with stationary states are self-sustained sources of radiation, fed by the pumping in the amplifying parts of the system.
(ii) That the energy of
these states is real means that the system is stabilized at the lasing
threshold.
(iii) When the system is leaky, the emitted radiation breaks
parity symmetry (i.e., the emission pattern is asymmetric). (iv) In the limit
of a non-leaking system, the emitted radiation intensity approaches a constant
value, and provides a direct measure of the non-Hermiticity of the system.
The internal energy density of radiation then diverges, which entails a
practical limitation for the implementation of ${\cal PT}$ symmetry in closed
systems.

These conclusions are obtained by employing the quantum-optical input-output formalism \cite{collet} in its scattering formulation \cite{gruner,beenakker,schomerus}. The scattering approach also provides insight into ${\cal PT}$ symmetry for classical waves \cite{cannata,ptscattering}, which defines the starting point of this Letter.

\begin{figure}
\includegraphics[width=\columnwidth]{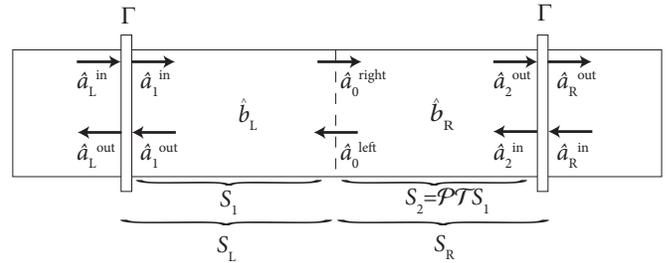}
\caption{Illustration of the scattering input-output approach to non-Hermitian ${\cal PT}$-symmetric systems, defining the scattering input-output operators $\hat a_{L,R}^{\rm in,\,out}$ and internal bosonic noise operators $\hat b_{L,R}$ in different parts of the system. The operators $\hat a_{0}^{\rm right,\,left}$ and $\hat a_{1,2}^{\rm in,\,out}$
feature in the intermediate steps of the calculation (for details see the text).
Semitransparent mirrors with transmission probability $\Gamma$ are introduced to study the limit $\Gamma\to 0$ of a closed (non-leaky) system. (In the context of present experiments with optical fibers  \cite{christodoulides}, this sketch relates to the transverse confinement.)} \label{fig1}
\end{figure}

{\em Scattering approach to non-Hermitian ${\cal PT}$-symmetric
systems}.---Probing the internal dynamics of an optical system by external
radiation naturally leads to the scattering scenario depicted in Fig.\
\ref{fig1}. The relation $a^{\rm out}=S a^{\rm in}$ between incoming and
outgoing wave amplitudes is provided by the scattering matrix \begin{equation}S(E)=\left(
\begin{array}{cc}
 r & t' \\
 t & r'
\end{array}
\right), \label{eq:smat}
\end{equation}
which contains blocks describing reflection ($r$, $r'$) and transmission ($t$, $t'$) when probed from the left or right, respectively. Each block consists of an $N\times N$-dimensional matrix, where $N$ is the number of modes at each entrance.  The scattering matrix is generally energy dependent (which
arises from the wavenumber dependence of constructive and destructive interference), and its poles determine the energies of quasibound states in the system \cite{smilansky}.

In general, the scattering matrix fulfills the following two reciprocity relations: The Onsager relation
\begin{equation}
S(\gamma,-B,E)=S^T(\gamma,B,E),
\end{equation}
and the relation
\begin{equation}
S(-\gamma,B,E)=[S^\dagger(\gamma,B,E^*)]^{-1}
\end{equation}
of classical microreversibility.
Here, $\gamma$ and $B$ characterize two possible sources of broken time-reversal symmetry:
absorption or amplification ($\gamma>0$ or $\gamma<0$), which contribute an imaginary symmetric (non-Hermitian) term to the Hamiltonian $H$, and magneto-optical effects ($B$), which contribute an imaginary antisymmetric (but still Hermitian) term.

Conventional time-reversal ${\cal T}H= H^*$ transforms solutions according to ${\cal T}\psi=\psi^*$, which interchanges incoming and outgoing states, and therefore transforms the scattering matrix according to
\begin{equation}
{\cal T} S(\gamma,B,E)=[S^*(\gamma,B,E)]^{-1}=S(-\gamma,-B,E^*).
\end{equation}
Assuming that energy is real, a system has ${\cal T}$ symmetry, ${\cal T}S=S$ hence $S^*=S^{-1}$,  if $S(\gamma,B)=S(-\gamma,-B)$, which requires $\gamma=B=0$ \cite{trsremark}.
 Parity ${\cal P}H(x)=H(-x)$ transforms solutions according to ${\cal P}\psi(x)=\psi(-x)$, which exchanges the left and right leads and  yields
\begin{equation}
{\cal P} S(\gamma,B,E)=\sigma_xS(\gamma,B,E)\sigma_x,
\end{equation}
where $\sigma_x$ is a Pauli matrix operating on the blocks in Eq.\ (\ref{eq:smat}).
The  ${\cal PT}$ operation on the scattering matrix is therefore given by
\begin{eqnarray}
{\cal PT} S(\gamma,B,E) &=& \sigma_x[S^*(\gamma,B,E)]^{-1}\sigma_x\\&=& \sigma_xS(-\gamma,-B,E^*)\sigma_x.
\end{eqnarray}

For Hermitian systems, ${\cal PT}$ symmetry implies $S=\sigma_x S^T\sigma_x$ \cite{baranger}. For  non-Hermitian systems, ${\cal PT}$ symmetry implies the additional condition ${\cal P} \gamma=-\gamma$ [$\gamma(x)=-\gamma(-x)$], i.e., there is a balance of absorption and amplification in parity-related regions.

Let us now explore from the scattering perspective how real-energy bound states appear in  ${\cal PT}$-symmetric systems.
As shown in Fig.\ \ref{fig1}, such systems can be constructed by joining two regions, where the
left region, with scattering matrix $S_1=
\left(
\begin{array}{cc}
 r_1 & t_1' \\
 t_1 & r_1'
\end{array}
\right)$, is
 ${\cal PT}$-symmetric to the right region,
$S_2={\cal PT}S_1$, which using standard block-inversion formulas can be written as
{\small\begin{equation}
S_2=\left(
\begin{array}{cc}\displaystyle
\frac{1}{({r_1'}-t_1r_1^{-1}{t_1'})^*} & \displaystyle(r_1'^{-1}t_1)^*\frac{1}{(t_1'r_1'^{-1}{t_1}-r_1)^*}\\[1cm]
\displaystyle(r_1^{-1}t_1')^*\frac{1}{(t_1r_1^{-1}{t_1'}-r_1')^*} & \displaystyle \frac{1}{(r_1-t_1'{r_1'}^{-1}t_1)^*}
\end{array}
\right).
\end{equation}}

Bound states can be studied by closing the system off by mirrors with small transmission probability $\Gamma\ll 1$,  described by a scattering matrix
\begin{equation}
S_\Gamma=-\left(
\begin{array}{cc}
\sqrt{1-\Gamma} & i\sqrt{\Gamma}\\[.3cm]
 i \sqrt{\Gamma} &\sqrt{1-\Gamma}
\end{array}
\right).
\end{equation}
The mirrors can be included by wave matching at their inward-facing faces, which amounts to an algebraic elimination of the amplitudes $a_{1,2}^{\rm in,\,out}$ in Fig.\ \ref{fig1}.

The scattering matrix of the left half of the system then reads
\begin{equation}\label{sl} \setlength{\arraycolsep}{10pt}
S_L=-\left(
\begin{array}{cc}\displaystyle
 \frac{r_1+\sqrt{1-\Gamma}}{1+r_1 \sqrt{1-\Gamma}} & \displaystyle\frac{i t_1' \sqrt{\Gamma} }{1+r_1 \sqrt{1-\Gamma}} \\[1cm]
 \displaystyle\frac{i t_1 \sqrt{\Gamma} }{1+r_1 \sqrt{1-\Gamma}} & \displaystyle\frac{t_1 t_1' \sqrt{1-\Gamma}}{1+r_1 \sqrt{1-\Gamma}}-r_1'
\end{array}
\right),
\end{equation}
while the scattering matrix $S_R={\cal PT} S_L$ of the right half again follows  from symmetry.
These scattering matrices relate amplitudes of ingoing and outgoing modes (defined in Fig.\ \ref{fig1}) according to
\begin{equation}
\left(\begin{array}{c}
 a_L^{\rm out}  \\
 a_0^{\rm right}
\end{array}\right)
=S_L^{\strut}
\left(\begin{array}{c}
 a_L^{\rm in}  \\
 a_0^{\rm left}
\end{array}\right)
,\quad
\left(\begin{array}{c}
 a_0^{\rm left} \\
 a_R^{\rm out}
\end{array}\right)
=S_R^{\strut}
\left(\begin{array}{c}
 a_0^{\rm right}  \\
 a_R^{\rm in}
\end{array}\right)
.\label{clscat}
\end{equation}

The scattering matrix of the composed system is obtained by algebraically eliminating the amplitudes
$a_0^{\rm left}$ and $a_0^{\rm right}$ at the interface between both regions.
For $\Gamma\to 0$, these amplitudes become singular when
\begin{equation} {\rm det}\,{\rm Im}\,(r'_L)={\rm det}\,\left[{\rm Im}\,\left(r_1'-\frac{t_1 t_1'}{1+r_1}\right)\right]=0,\label{qcond}
\end{equation}
which (due to the general connection of scattering poles and quasibound states) is the quantization condition of the closed system. Determinantal quantization conditions of this kind have been introduced for general systems in Ref.\ \cite{smilansky}; they also form the basis of exact numerical techniques as reviewed in \cite{baecker}.
The ${\cal PT}$-specific  version (\ref{qcond}) of the quantization condition requires that the $N$ real column vectors of ${\rm Im}\,(r'_L)$
be linearly dependent, which generically can be achieved by varying a single real parameter such as $E$
(identifying this as a problem of codimension one). Therefore, we recover that closed ${\cal PT}$-symmetric
systems typically possess a number of bound states with real energy, even though the Hamiltonian is not Hermitian.
Condition (\ref{qcond}) also admits complex eigenvalues, which then appear in complex-conjugated pairs.

{\em Quantum noise}.---The scattering approach can be extended to include quantum noise by passing from wave amplitudes $a^{\rm in}$, $a^{\rm out}$ to bosonic annihilation operators
$\hat a^{\rm in}$, $\hat a^{\rm out}$, respectively. This defines the scattering variant of the input-output formalism \cite{collet,gruner,beenakker,schomerus}, which has been used
to describe systems that are exclusively absorbing or amplifying.
To adapt the approach to ${\cal PT}$-symmetric systems, where both effects are combined, we formally separate the absorbing regions from the amplifying regions, and then join them together similar to the description of classical waves, given above.

For definiteness let us assume that the left half of the system is purely absorbing.
For this part, the input-output scattering relations then take the form
\begin{equation}
\left(\begin{array}{c}
 \hat a_L^{\rm out}  \\
 \hat a_0^{\rm right}
\end{array}\right)=S_L\left(\begin{array}{c}
 \hat a_L^{\rm in}  \\
 \hat a_0^{\rm left}
\end{array}\right)+Q_L{\hat b}_L,
\label{inout1}
\end{equation}
which connects the ingoing and outgoing modes to
bosonic operators $\hat b_L$ and  $\hat b_R$ representing quantum fluctuations in the left and right part of the medium. These operators can be defined microscopically
in the framework of system-and-bath approaches, or phenomenologically as  operator-valued Langevin forces  \cite{collet,gruner}. The properties of these operators can be determined from the condition that
both $\hat a^{\rm in}$ and $\hat a^{\rm out}$ have to satisfy standard canonical commutation relations. This dictates that the coupling matrix $Q_L$ satisfies the fluctuation-dissipation theorem\cite{beenakker}
\begin{equation}
Q_L^{}Q_L^\dagger=1-S_L^{}S_L^\dagger.\label{eq:use1}
\end{equation}
 In the right half of the system, where the medium is amplifying, we have
\begin{equation}
\left(\begin{array}{c}
  \hat a_0^{\rm left}  \\
 \hat a_R^{\rm out}
\end{array}\right)=S_R\left(\begin{array}{c}
 \hat a_0^{\rm right} \\
\hat a_R^{\rm in}
\end{array}\right)+Q_R^{}\hat b_R^\dagger,
\label{inout2}
\end{equation}
where the commutation relations now dictate coupling  to creation operators, with \begin{equation}Q_R^{}Q_R^\dagger=S_R^{}S_R^\dagger-1. \label{eq:use2}\end{equation}
By assumption, the operators $\hat b_L^\dagger$ and $\hat b_R$ commute with ${\hat a}^{\rm in}$; however,
according to Eqs.\ (\ref{inout1}) and (\ref{inout2}) they do not commute with ${\hat a}^{\rm out}$, which is a manifestation of broken micro-reversibility in quantum optics.

We can now describe the full ${\cal PT}$-symmetric system  by algebraically
eliminating the interface operators $\hat a_0^{\rm left}$ and  $\hat a_0^{\rm
right}$. In the absence of incoming radiation, the intensity emitted to the
left and right is then given by
\begin{equation}I_L(E)=\frac{1}{2\pi}\langle {\hat a}^{\rm
out \dagger}_L {\hat a}^{\rm out}_L\rangle, \quad I_R(E)=\frac{1}{2\pi}\langle
{\hat a}^{\rm out \dagger}_R {\hat a}^{\rm out}_R\rangle,
\end{equation}
 which can be
evaluated assuming
\begin{equation}
\langle {\hat b}^{\dagger}_L {\hat b}_L^{}\rangle=0,\quad \langle {\hat
b}^{\dagger}_R {\hat b}_R^{}\rangle=0 \label{eq:use3}
\end{equation}
(ground-state population in the absorbing regions and total population inversion in the
amplifying regions; these conditions minimize the quantum noise).

Let us first consider the case of a single-mode resonator ($N=1$) with  purely ballistic
internal dynamics and absorption in the left region \cite{znojil}, with scattering matrices
\begin{equation}\label{res1}
S_1=\left(
\begin{array}{cc}
 0 & t_1 \\
 t_1 & 0
\end{array}
\right),\quad
S_2=\left(
\begin{array}{cc}
 0 & 1/t_1^* \\
  1/t_1^* & 0
\end{array}
\right),
\end{equation}
where $|t_1|<1$.
Including the mirrors, the
total scattering matrix is \cite{remark4}
\begin{equation}\label{res3}
S=\left(
\begin{array}{cc}\setlength{\arraycolsep}{10pt}\displaystyle
\frac{\sqrt{1-\Gamma}  \left({t_1^*}^2-t_1^2\right)}{t_1^2 (1-\Gamma)-{t_1^*}^2} & \displaystyle\frac{|t_1|^2 \Gamma }{t_1^2 (1-\Gamma)-{t_1^*}^2} \\[.8cm]
\displaystyle \frac{|t_1|^2\Gamma}{t_1^2 (1-\Gamma)-{t_1^*}^2} & \displaystyle\frac{\sqrt{1-\Gamma}  \left({t_1^*}^2-t_1^2\right)}{t_1^2 (1-\Gamma)-{t_1^*}^2}
\end{array}
\right),
\end{equation}
and the quantization condition (\ref{qcond}) for the closed resonator takes the form ${\rm Im}\, t_1^2=0$ (indeed, the scattering matrix elements diverge under this condition and $\Gamma\to 0$).
Following the quantum-optical procedure described above [and using, in particular, Eqs.\ (\ref{eq:use1}), (\ref{eq:use2}), and (\ref{eq:use3})],
we find that this resonator emits radiation of intensity
\begin{eqnarray}&&I_L(E)=
\frac{ \Gamma \left(|t_1|^{-2}-1\right)\left(1-\Gamma +|t_1|^2\right)}{2\pi|(t_1/t_1^*)^2-1+\Gamma|^2},
\\&&I_R(E)=\frac{ \Gamma \left(1-|t_1|^{2}\right) \left(1-\Gamma +|t_1|^{-2}\right)}{2\pi|(t_1/t_1^*)^2-1+\Gamma|^2}.
\end{eqnarray}
Since $|t_1|<1$ this gives $I_R>I_L$, the difference being
\begin{equation}\Delta I(E)=I_R(E)-I_L(E)=\frac{\Gamma ^2 \left(|t_1|^{-1}-|t_1|\right)^2}{2\pi|(t_1/t_1^*)^2-1+\Gamma|^2}.
\end{equation}
Therefore, the emission from the right exit, close to the amplifying region of the medium, is larger than the emission from the left exit, close to the absorbing region of the medium.
The overall  output intensity to both sides can be written as
\begin{equation}I(E)=I_L(E)+I_R(E)=\frac{\Gamma (2-\Gamma)\left(|t_1|^{-2}-|t_1|^2\right)}{2\pi|(t_1/t_1^*)^2-1+\Gamma|^2}.
\end{equation}

Close to quantization in the closed system [$\Gamma\ll 1$,  $E\approx E_0$, where $E_0$ fulfills the quantization condition ${\rm Im}\, t_1^2(E_0)= 0$], the emission pattern becomes symmetric (as a consequence of ${\cal PT}$ invariance) and approaches a Lorentzian of the form
\begin{equation}I_L(E)=I_R(E)=\frac{\Gamma \left(|t_0|^{-2}-|t_0|^2\right)}{2\pi|2i\tau (E-E_0)+\Gamma|^2}.
\end{equation}
Here $t_0= t_1(E_0)$, while
\begin{equation}
\tau=2\,{\rm Im}\,\frac{1}{t_1}\left.\frac{dt_1}{dE}\right|_{E=E_0}\approx 2L/c
\end{equation}
is the transmission delay time of propagation between the two mirrors (with $L$ the length of each region and $c$ the speed of light) .
The full width at half maximum is given by $\Delta E=\Gamma/\tau$. While this width shrinks to zero as the system is closed off, remarkably the total intensity
\begin{equation}I_{\rm tot}=\int I(E)\, dE= \frac{|t_0|^{-2}-|t_0|^2}{2\tau}
\end{equation}
remains finite, and can be interpreted as a direct measure of the degree of non-Hermiticity of the system (for ballistic transport, Hermiticity implies  $|t_0|= 1$).

In the case of a single-mode resonator with backscattering (where $r_1$ and $r_1'$ are finite), compact expressions can still be obtained
as long as the leakage remains small ($\Gamma\ll 1$), implying according to  Eq.\ (\ref{sl}) that $|r_L+1|,|t_L|, |t_L'|\ll 1$. The emission pattern is then still symmetric, with intensity
\begin{equation}
I_L(E)=I_R(E)=\frac{1}{2\pi}\frac{(1-|r_L'|^2)|t_L'|^2}{|2(\im r_L')-it_Lt_L'|^2}.
\end{equation}
Linearization around the quantization condition again reveals a Lorentzian line shape,
with line width \begin{equation}
\Delta E=\,{\rm Re}\, \left(\frac{d[({\rm Im}\, r_L')/t_Lt_L']}{dE }\right)^{-1}.
\end{equation}
 Accounting for the scaling (\ref{sl}) of scattering coefficients with $\Gamma$,
the total intensity $I_{\rm tot}\propto (1-|r_L'|^2)$ again remains finite as $\Gamma\to 0$.
In the Hermitian case, this limit would imply  $|r_L'|= 1$, so that the intensity vanishes. Therefore, the emitted radiation is still a direct measure of the degree of non-Hermiticity.

Following the general formalism described above, the observations for one-dimensional scattering  can be extended to the  general case of ${\cal PT}$-symmetric systems with many modes, for which compact expressions are no longer available. The emitted intensity generally remains finite even in the limit of a closed system. Because the expectation values $\langle \hat a_0^{\rm left\dagger}\hat a_0^{\rm left}\rangle \propto\Gamma^{-1}$, $\langle \hat a_0^{\rm right\dagger}\hat a_0^{\rm right}\rangle \propto\Gamma^{-1}$ of the internal operators formally diverge in this limit, this is accompanied by a diverging internal energy density, which can be interpreted as the source of this radiation.
Deviations from perfect ${\cal PT}$ symmetry can be
incorporated into the scattering formalism by taking $S_R \neq {\cal
PT} S_L$, which results in an asymmetric emission
pattern even in the limit of non-leaky mirrors.

In standard laser theory \cite{siegman}, an active optical
medium is below threshold (and stable) as long as all fundamental modes
are damped, which is characterized by complex energies with a
negative imaginary part. With increasing pumping, the loss of the modes
within the amplification window is counteracted by gain, and as soon as
one of the energies  acquires a positive imaginary
part the system becomes instable: the intensity
increases exponentially until saturation sets in. Therefore,
the radiation features described above are characteristic of a laser stabilized at threshold, the threshold condition being that the energy of the lasing mode is real. The system then is marginally stable, and the internal radiation energy accumulates due to the quantum fluctuations.
In practice, this leads to saturation in the amplifying parts of the system and therefore identifies an obstacle for the implementation of strict ${\cal PT}$ symmetry in closed optical systems.
When the system is slightly opened up, the threshold condition is no longer met and the internal energy density is finite, while the emitted intensity remains a direct measure of the non-Hermiticity of the system.

{\em Concluding remarks.}---In summary, the present Letter demonstrates that accounting for quantum noise, optical realizations of non-Hermitian ${\cal PT}$-symmetric systems emit radiation of an intensity that provides a direct measure of non-Hermiticity.

This has both practical as well as fundamental implications.
The fundamental consequences arise because non-Hermitian ${\cal PT}$-symmetric  systems with an entirely real spectrum define a consistent unitary theory of quantum mechanics \cite{bender2002,mostafazadeh}, formalized by the concept of quasi-Hermiticity, which introduces a new scalar product based on a generalized conjugation operation ${\cal C}$ that satisfies ${\cal C}^2=1$, $[{\cal C},H]=[{\cal C},{\cal PT}]=0$. The self-sustained radiation identified here shows that
accounting for quantum noise, non-Hermitian ${\cal PT}$-symmetric systems are physically distinct from ordinary Hermitian quantum systems: the canonical commutation relations for the input and output operators are only invariant under unitary transformations, which constraints the possibility to introduce alternative scalar products. This enforces classical distinctions based on the transparency of such systems \cite{ptscattering}.

From a practical perspective, the self-sustained radiation can be used as an indicator of successfully implemented non-Hermitian ${\cal PT}$ symmetry in leaky  systems, while the accompanying marginal instability and diverging internal energy density signifies a practical obstacle for its implementation in the limit of no leakage. Furthermore, these findings identify a hitherto unexplored arena to study quantum fluctuations in active systems close to the lasing threshold.

Non-optical realizations of ${\cal PT}$-symmetric systems offer a wide range of connections to quantum field theories \cite{bender2007a}, and while the results in the present paper are not directly transferable in detail, they suggest that quantum noise should provide fundamental insight into such systems, as well.


\begin{thebibliography}{99}



\bibitem{bender1998}
C. M. Bender and S. Boettcher, Phys. Rev. Lett. \textbf{80}, 5243
(1998).


\bibitem{bender2002}
C. M. Bender, D. C. Brody, and H. F. Jones,  Phys. Rev. Lett. \textbf{89}, 270401
(2002);
Am. J. Phys. \textbf{71}, 1095 (2003).


\bibitem{mostafazadeh}
F. G. Scholtz, H. B. Geyer, and F. J. W. Hahne, Ann. Phys. (NY) \textbf{213}, 74 (1992);
A. Mostafazadeh, J. Math. Phys. (N.Y.) \textbf{43},
205 (2002);
{\em ibid.}  2814 (2002);
{\em ibid.} 3944 (2002); J. Phys. A \textbf{36},  7081 (2003);
preprint arxiv:0810.5643.


\bibitem{bender2007a}
C. M. Bender, Rep. Prog. Phys. \textbf{70}, 947 (2007).

\bibitem{christodoulides}
C. E. R{\"u}ter \emph{et al.},
Nature Phys. \textbf{6}, 192 (2010);
Z. H. Musslimani, K. G. Makris, R. El-Ganainy, and D. N.  Christodoulides,
Phys. Rev. Lett. \textbf{100},  030402   (2008);
K. G. Makris, R. El-Ganainy,  D. N.  Christodoulides, and Z. H. Musslimani,
{\em ibid.} \textbf{100}, 103904  (2008); A. Guo {\em et al.}, {\em ibid.} \textbf{103},  093902   (2009).


\bibitem{einstein}
R. Loudon, \emph{The Quantum Theory of Light}, 3rd ed. (Oxford University Press, New York, 2001).


\bibitem{collet} M. J. Collett and C. W. Gardiner, Phys. Rev. A  \textbf{30}, 1386 (1984);
C. W. Gardiner and M. J. Collett, Phys. Rev. A \textbf{31}, 3761 (1985).

\bibitem{gruner}
T. Gruner and D.-G. Welsch, Phys. Rev. A \textbf{54}, 1661 (1996).

\bibitem{beenakker}C. W. J. Beenakker, Phys. Rev. Lett. \textbf{81}, 1829
    (1998).

\bibitem{schomerus} H. Schomerus, K. Frahm, M. Patra, and C. W. J. Beenakker,
    Physica A \textbf{278}, 469 (2000).

\bibitem{cannata} For ${\cal PT}$-symmetric scattering theory in one dimension see F. Cannata, J. P. Dedonder
and A. Ventura, Ann. Phys. \textbf{322}, 397 (2007).

\bibitem{ptscattering} M. V. Berry, J. Phys A \textbf{41}, 244007 (2008);
H. F. Jones, Phys. Rev. D \textbf{76}, 125003 (2007); \emph{ibid.}
\textbf{78}, 065032 (2008).

\bibitem{trsremark}
In Hermitian systems ($\gamma=0$), ${\cal T}$ is equivalent to
the alternative time-reversal ${\cal T}'H= H^T$, for which
${\cal T}' S(\gamma,B,E)=S^T(\gamma,B,E)=S(\gamma,-B,E)$.
For ${\cal T}'$ symmetry,  $S=S^T$, hence $S(\gamma,B)=S(\gamma,-B)$, which requires $B=0$, while $\gamma$ can be finite. While absorption or amplification generally breaks ${\cal T}$ symmetry,
${\cal T}'$ symmetry is therefore still observed in  optical systems which show no magneto-optical effects, even when they are absorbing or amplifying
(in that case $H=H^T$ is non-Hermitian but symmetric).
A system with ${\cal T}{\cal T}'$ symmetry obeys
$S^{-1}=S^\dagger$, hence $S(-\gamma,B)=S(\gamma,B)$, requiring $\gamma=0$ so that the Hamiltonian is Hermitian.

\bibitem{smilansky}E. Doron and U. Smilansky, Phys. Rev. Lett. \textbf{68}, 1255 (1992).



\bibitem{baranger}
H. U. Baranger and P. A. Mello, Phys. Rev. B \textbf{54}, R14297 (1996);
V. A. Gopar, S. Rotter, and H. Schomerus, Phys. Rev. B \textbf{73}, 165308 (2006);
M. Kopp, H. Schomerus, and S. Rotter, Phys. Rev. B \textbf{78}, 075312 (2008); R. S. Whitney, H. Schomerus, and M. Kopp,
Phys. Rev. E \textbf{80}, 056209 (2009); \emph{ibid.}  \textbf{80}, 056210 (2009).

\bibitem{baecker} A. B{\"a}cker, in \emph{The Mathematical Aspects of Quantum Maps},
edited by M. D. Esposito and S. Graffi, Lecture Notes in Physics
Vol. 618 (Springer, Berlin, 2003), p. 91. 


\bibitem{znojil}
M. Znojil, Phys. Lett. A \textbf{285}, 7 (2001).

\bibitem{remark4}
The
left and right halves of the resonator are described by
\begin{eqnarray*}\label{res2}
&&S_{L}=-\left(
\begin{array}{cc}\setlength{\arraycolsep}{10pt}
\sqrt{1-\Gamma}  & i t_1 \sqrt{\Gamma} \\[.5cm]
 i t_1 \sqrt{\Gamma} & t_1^2 \sqrt{1-\Gamma}
\end{array}
\right),
\\[.5cm]
&&S_{R}=-\left(
\begin{array}{cc}\setlength{\arraycolsep}{10pt}\displaystyle
\frac{\sqrt{1-\Gamma}}{{t_1^*}^2}  & \displaystyle\frac{i  \sqrt{\Gamma}}{t_1^*} \\[.5cm]
\displaystyle\frac{i  \sqrt{\Gamma}}{t_1^*} & \sqrt{1-\Gamma}
\end{array}
\right).
\end{eqnarray*}
\vspace*{.3cm}
\bibitem{siegman} A. E. Siegmann, \emph{Lasers} (University Science Books, Mill Valley,
CA, 1986).



\end{thebibliography}
\end{document}